\documentclass{ieeeaccess}

\usepackage{cite}
\usepackage{amsmath,amssymb,amsfonts}
\usepackage{algorithmic}
\usepackage{graphicx}
\usepackage{textcomp}

\usepackage[utf8]{inputenc}
\usepackage{amsfonts}
\usepackage{nicefrac}
\usepackage{microtype}
\usepackage{soul}
\usepackage{color}
\usepackage{amsmath}
\usepackage{graphicx}
\usepackage{subfigure}
\usepackage{algorithm}
\usepackage{algorithmic}
\usepackage{url}
\usepackage{balance}

\usepackage{dblfloatfix}    

\usepackage[table,xcdraw]{xcolor}

\soulregister\cite7

\def\BibTeX{{\rm B\kern-.05em{\sc i\kern-.025em b}\kern-.08em
    T\kern-.1667em\lower.7ex\hbox{E}\kern-.125emX}}
\begin{document}

\title{JSDoop and TensorFlow.js: Volunteer Distributed Web Browser-Based Neural Network Training}

\author{\uppercase{Jos\'{e} \'{A}. Morell}\authorrefmark{1},
\uppercase{Andr\'{e}s Camero\authorrefmark{1}, and Enrique Alba}.\authorrefmark{1}}
\address[1]{Departamento de Lenguajes y Ciencias de la Computaci\'{o}n, Universidad de M\'{a}laga, Andaluc\'{i}a Tech, Espa\~{n}a (e-mail: jamorell@lcc.uma.es, andrescamero@uma.es, eat@lcc.uma.es)}
\tfootnote{This research was partially funded by Ministerio de Econom\'ia, Industria y Competitividad, Gobierno de Espa\~na, and European Regional Development Fund grant numbers TIN2017-88213-R (\url{http://6city.lcc.uma.es}), RTC-2017-6714-5 (\url{http://ecoiot.lcc.uma.es}), and UMA18-FEDERJA-003 (PRECOG).
Universidad de M\'alaga, Campus Internacional de Excelencia Andaluc\'ia TECH.
Jos\'{e} \'{A}. Morell is supported by an FPU grant (FPU16/02595) from the Ministerio de Educaci\'on, Cultura y Deporte, Gobierno de Espa\~na.}

\corresp{Corresponding author: Jos\'{e} \'{A}. Morell (e-mail: jamorell@lcc.uma.es).}

\begin{abstract}
In 2019, around 57\% of the population of the world has broadband access to the Internet. Moreover, there are 5.9 billion mobile broadband subscriptions, i.e., 1.3 subscriptions per user. 
So there is an enormous interconnected computational power held by users all around the world.
Also, it is estimated that Internet users spend more than six and a half hours online every day. But in spite of being a great amount of time, those resources are idle most of the day. Therefore, taking advantage of them presents an interesting opportunity. In this study, we introduce JSDoop, a prototype implementation to profit from this opportunity. In particular, we propose a volunteer web browser-based high-performance computing library. 
JSdoop divides a problem into tasks and uses different queues to distribute the computation. 
Then, volunteers access the web page of the problem and start processing the tasks in their web browsers. 
We conducted a proof-of-concept using our proposal and TensorFlow.js to train a recurrent neural network that predicts text. We tested it in a computer cluster and with up to 32 volunteers. 
The experimental results show that training a neural network in distributed web browsers is feasible and accurate, has a high scalability, and it is an interesting area for research.
\end{abstract}

\begin{keywords}
Artificial Intelligence, Browsers, Collaborative Work, Distributed Algorithms, Distributed Computing, Internet, JavaScript, JSDoop, Neural networks, TensorFlow, Volunteer Computing
\end{keywords}

\titlepgskip=-15pt

\maketitle

\section{Introduction}
\vspace{0.025em}
While the population of the World in 2017 is estimated to 7.6 billion~\cite{desa2017world}, the number of Internet users is estimated to be nearly 4.4 billion, and in 2018, it has grown more than 9\%~\cite{digital2019report}. In other words, Internet has 57\% penetration globally. Moreover, according to Ericsson, there are 5.9 billion mobile broadband subscriptions around the world~\cite{ericsson2019report} (i.e., 1.3 subscriptions per user), and more than 30\% of the population has a fixed broadband access~\cite{oecd2015digital}.

\vspace{0.6em}

On the other hand, Internet users spend more than 6.5 hours online each day on average~\cite{digital2019report}, but in spite of the \emph{great amount} of time, it seems reasonable to think that most of the time these interconnected devices are idle. Therefore, finding a way to take advantage of this unused processing capacity represents an interesting opportunity.

\vspace{1em}

This huge (idle) computational power would be very useful for science and society. However, taking advantage of it is challenging. First, we need to get access to those billion devices held by users all around the world, and second (but not least), we need to find a way of managing the globally distributed resources transparently. In this sense, several attempts have been made, and amid them, \emph{volunteer computing} (VC) arises as a prominent approach~\cite{anderson2019boinc,cerin2012desktop,chorazyk2017lightweight}.
\vspace{0.5em}

VC is a type of distributed computing in which people (called \textit{volunteers}) voluntarily donate their computing resources to a project. In spite of its success, VC still imposes many challenges. In particular, these systems are composed of a wide variety of heterogeneous hardware with different performances. Therefore, they have to deal with asynchronism, fault tolerance, and with devices that join and leave the system arbitrarily.

It is no surprise that dedicated grid and cloud platforms are more efficient than heterogeneous hardware collaborating over the Internet (i.e., VC). However, in contrast to such dedicated approaches, VC provides researchers with a computing power that is not attainable otherwise, at a lower cost, and reducing energy consumption (we are using idle resources instead of powering up new ones). 

Also, VC encourages public interest in science and provides people with a voice in determining the directions of scientific research. If the data or target application has a clear public vocation, many users will share their resources for humanitarian purposes, such as studying cancer, water quality in a city, or helping to reduce energy consumption. All these reasons have motivated several scientific projects that accomplish their mission thanks to volunteers, e.g., IBM World Community Grid~\cite{ibm2019communitygrid}, Rosetta@home~\cite{das2007structure}, SETI@home~\cite{korpela2001seti}, among others~\cite{anderson2019boinc}.

In the last decade, thanks to the notorious improvement in the processing capacity of web browsers~\cite{tilkov2010node}, and the release of powerful software libraries for them (e.g., WebGl, and TensorFlow.js), the browser-based volunteer computing (BBVC) has gained popularity~\cite{wilkinson2014qmachine,meeds2015mlitb,chorazyk2017lightweight}. BBVC offers portability, flexibility, and ubiquity \emph{out of the box}. However, due to the rapid development of all these new technologies, there are yet few attempts to combine all these improvements to bring high-performance computing (HPC) to a BBVC platform.

With this in mind, we set the main objective of this study: to propose an HPC BBVC library. Therefore, the main results and contributions of this work are as follows:

\Figure[!h](topskip=0pt, botskip=0pt, midskip=0pt)[width=\textwidth]{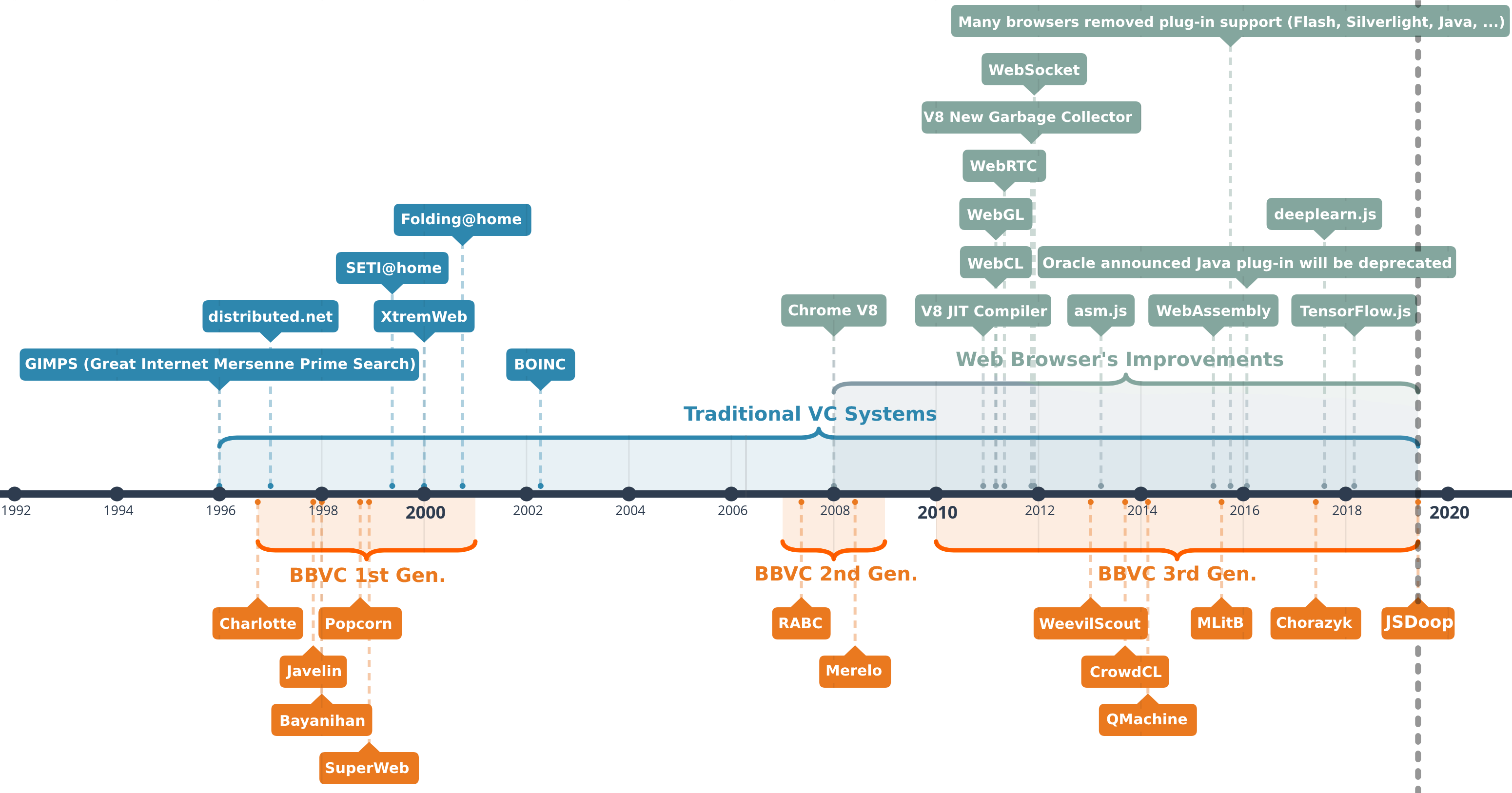}
{The evolution of volunteer computing and browser-based volunteer computing.\label{fig:generations_timeline}}

\begin{itemize}
    \item \emph{High-performance BBVC}: We introduce JSDoop, an HPC BBVC library. This open source library will allow developers to include background HPC tasks on their web applications, without interrupting the user experience on the site, and without installing additional software. Moreover, the user will continue using the web browser as usual while JSDoop is running.
    
    \item \emph{Web browser-based distributed neural network training}:  We conducted a proof-of-concept to show that distributed training of neural networks in the browser is possible (using TensorFlow.js). Specifically, we trained a recurrent neural network (RNN) to predict text~\cite{mckeown1992text} in distributed browsers using TensorFlow.js and JSDoop. The results show that it is feasible, as well as scalable, and it is also an interesting area to explore.
\end{itemize}

The remainder of this study is organized as follows: the next section briefly presents VC and BBVC related works. Section~\ref{sec:distributed_deeplearning} discusses distributed deep learning in the web browser. Section~\ref{sec:jsdoop} introduces our proposal, JSDoop. Section~\ref{sec:experiments} shows the experimental study and the results. Section~\ref{sec:validity} discusses the threats to validity. And finally, Section~\ref{sec:conclusions} outlines the main conclusions and proposes future work.

\section{Volunteer Computing and Browser-Based Volunteer Computing}\label{sec:vc-bbvc}

This section briefly reviews the development of VC and BBVC: (\emph{A}) we introduce VC systems; (\emph{B}) we present the first and second generation of BBVC; (\emph{C}) we outline the third generation of BBVC; (\emph{D}) we present the desirable features of current VC systems; (\emph{E}) we define the requirements for JSDoop in terms of the desirable features for VC systems.

Figure~\ref{fig:generations_timeline} summarizes the evolution of VC and BBVC. Also, it highlights the most important technological improvements made in the last decade to web browsers, as well as relevant works in the field.

\subsection{Volunteer Computing}\label{sec:trad-vc-systems}

The first reminiscences of VC are dated back to the mid-1990s. The original idea was to use volatile idle resources spread over the world to solve complex problems. Specifically, in 1996, the project GIMPS (Great Internet Mersenne Prime Search, \url{http://www.mersenne.org)}) released the first known VC platform. Its aim was (and is, it is running still) to search for Mersenne prime numbers, and to encourage volunteers, they offered a \$50,000 reward for the volunteer who discovered a prime number with at least 100,000,000 decimal digits.

A bit later, in 1997, the \emph{distributed.net} project emerged as an effort to break the RC5-56 portion of the RSA Secret-Key Challenge. In 1999, the University of Berkeley started the SETI@home project~\cite{korpela2001seti}, which they later released as a platform for general volunteer and grid computing, BOINC (Berkeley Open Infrastructure for Network Computing)~\cite{anderson2004boinc}. That year also, the Folding@home project (\url{https://foldingathome.org/}) was launched for disease research that simulates protein folding, computational drug design, and other types of molecular dynamics. A year later, XtremeWeb~\cite{cerin2012desktop,fedak2001xtremweb} appeared as an open source software to build a lightweight desktop grid.

VC systems have proved to be especially well suited for CPU-intensive applications that can be broken into many independent and autonomous tasks but are somewhat inappropriate for data-intensive tasks. Due to its centralized nature, VC requires all data to be served by a group of centrally maintained servers. Despite the improvements made so far, such as dedicated protocols or cycle stealing during idle CPU time, their lack of accessibility and usability are still significant drawbacks~\cite{cushing2013distributed}.

Each VC platform has different ways of attracting volunteers~\cite{sarmenta1998bayanihan,shahri2014gamification}. They might be true volunteers (i.e., altruists), paid volunteers, forced volunteers, or even use \emph{gamification} to attract volunteers. However, no matter their origin, VC systems require their users to run specific software, i.e., installation is required, and it is well-known that some people do not want (or know how) to install unfamiliar software on their machines~\cite{fabisiak2017browser}. They are not sure what they are installing on their devices and may even think that the application has access to their privacy~\cite{cusack2006volunteer}.

In spite of these difficulties, VC has grown to achieve unprecedented processing capacity. One such example is BOINC, that has 27 PetaFLOPS average computing power available, more than 300 thousand active participants and nearly 850 thousand active computers~\cite{anderson2019boinc}. As a remark, Summit or OLCF-4 (among the fastest supercomputers in the world) achieve 143.5 PetaFLOPS in the LINPACK benchmark (\url{https://www.top500.org/lists/2018/11/}). It is amazing that a volunteer computing platform can even perform close to that of the world’s fastest supercomputer.

\subsection{BBVC 1st and 2nd Generation}\label{sec:bbvc12}

At the very beginning of VC, some authors started to explore the use of incipient web browser to coordinate distributed computation efforts, giving birth to BBVC. Their initial approach was to implement Java applets that ran in the browser (e.g., Charlotte~\cite{kedem1996charlotte}, Javelin~\cite{christiansen1997javelin}, Bayanihan~\cite{sarmenta1998bayanihan}, Popcorn~\cite{camiel1997popcorn}, SuperWeb~\cite{alexandrov1997superweb}, among others). In particular, the user had to click on a link to download an applet and then allow it to run. As a remark from the expectations generated, some authors predicted in 1997 that by 2007 supercomputing in the web browser would be possible~\cite{fox1997petaops}. One of the disadvantages of this generation was that a user interaction was needed to start the application. It was also necessary to install a Java plugin, which was very slow to start. 

Nowadays, Java and Flash plugins have become obsolete due to serious security problems. Moreover, from 2015, many browser vendors have started to removed plugin support, eliminating the ability to embed Flash, Silverlight, Java and other plugin based technologies. 

BBVC was forgotten until 2007 when a second generation based on JavaScript appeared. Notice that by that time Javascript did not have modern features (e.g., JIT compiler, WebSocket, etc.). Among the notable works, we might highlight \emph{RABC}, by Konishi et al.~\cite{konishi2007rabc}, that proposes a large scale distributed system based on AJAX, and the one of Merelo et al.~\cite{merelo2008asynchronous}, that implements a distributed evolutionary algorithm in a P2P and volunteer computing environment. 

The major advantage of the second generation was that no user interaction was required. The program started simply by visiting a website. Despite this important improvement, they still presented a significant issue: performance. In 2007, JavaScript was between 9.8 to 23.2 times slower than Java~\cite{konishi2007rabc}, and between 20 to 200 times slower than C~\cite{klein2007unwitting}. The low performance, summed to the lack of support for multi-thread and the inability for direct communications between browsers, made second-generation BBVC failed.

\subsection{BBVC 3rd Generation}\label{sec:bbvc3}

In 2008, the great leap in computing capacity in the browser came~\cite{haas2017bringing}. In particular, Google released Chrome V8 (\url{https://v8.dev/}), a high-performance JavaScript and WebAssembly engine. But it is really in 2010 when Google released a new compiling infrastructure named Crankshaft and WebWorker was announced (multi-thread Javascript, \url{https://html.spec.whatwg.org/multipage/workers.html}), that the browser experimented a major change towards a competitive performance.

Later, in 2011, there were launched two important technologies for HPC in the browser: WebCL (\url{https://www.khronos.org/webcl/}) and WebGL (\url{https://www.khronos.org/webgl/}).
WebCL is a JavaScript link to OpenCL, a heterogeneous parallel computing framework that leverages CPUs and multicore GPUs within the web browser without the use of plugins. Some works have already used WebCL (e.g., WeevilScout~\cite{cushing2013distributed}, CrowdCL~\cite{macwilliam2013crowdcl}). However, WebCL is still under development, and no browser supports it natively. Thus, the only way to use it is through browser extensions. 

On the other hand, WebGL allows GPU-accelerated usage of physics and image processing in the web browser without the use of plugins. Although WebGL was originally developed for graphics rendering, it is also used for other applications like machine learning (ML) in the browser (e.g., TensorFlow.js, \url{https://github.com/tensorflow/tfjs}).

Also, two important communication technologies were launched in 2011: WebSocket and WebRTC. WebSocket (\url{https://www.w3.org/TR/websockets/}) provides full duplex communication channels over a single TCP connection. This technology is more suitable than HTTP in situations where low latency is required~\cite{morell2017distributed}. WebRTC~\cite{bergkvist2012webrtc,dias2015browsercloud} is a browser-based real-time peer-to-peer communication without plugins.

To allow native code (e.g., C, C++) to run in the web browser, Mozilla released  \emph{asm.js} in 2013 (\url{http://asmjs.org}). A few years later, in 2015, WebAssembly (WASM)~\cite{haas2017bringing} move forward in the same direction. 
Both approaches use a source-to-source compiler (e.g., Emscripten, \url{https://emscripten.org}) to translate the original source code to the desired format (i.e., asm.js or WASM code). There are many successful examples of migrating native desktop applications to the web using these technologies. For instance, SQLite (\url{https://www.sqlite.org}), Unreal Engine 3 (\url{https://www.unrealengine.com}), and AutoCAD (\url{https://web.autocad.com}).

Many important security improvements to web browsers have been made during this BBVC generation. Maybe the most important one (for BBVC) is the introduction of the sandbox~\cite{hunt2018ryoan}. In modern web browsers, web applications run isolated, i.e., they can only access a limited set of resources. As a consequence, users can run applications without installing new software on their devices, and they can be sure that those applications will not access their privacy, as long as they do not have security vulnerabilities.

All these new technologies have enabled a brand new potential. Thus, many interesting BBVC platforms have been released so far. For example, QMachine~\cite{wilkinson2014qmachine} and the one proposed by Chorazyk et al.~\cite{chorazyk2017lightweight}.

There are also some BBVC ML frameworks (e.g., MLitB~\cite{meeds2015mlitb} and OpenML~\cite{vanschoren2014openml}). However, until today, none of them have taken advantage of all the available technologies (i.e., WebGL, WebCL, WASM, etc.). Even more, TensorFlow.js~\cite{smilkov2019tensorflow}, the implementation of the popular ML framework of Google in Javascript, was released in 2018. Therefore, there is almost no evidence of the integration of Tensorflow.js in BBVC systems. Note that Google previously released \emph{deeplearn.js}, but now it is integrated into the core of TensorFlow.js.

Thanks to all these improvements, current BBVC systems stand out for their portability, extraordinary computing power potential and for being more secure than desktop applications. Therefore, we propose to integrate TensorFlow.js, WASM, and WebGL into a BBVC system. Note that we do not propose to include WebCL because it is still work-in-progress.

\subsection{Desirable Features for Volunteer Computing Platforms}

Designing and implementing a VC system is not simple. One has to deal with many challenges, and there are so many technologies available. To ease this process, some authors have proposed to define a set of desirable features that a VC system should have. 

In particular, we propose to use as a guideline the list described by Fabisiak and Danilecki~\cite{fabisiak2017browser}. Table~\ref{tab:desirable_features} summarizes the desired features.

\begin{table}[h]
	\caption{Desirable Features for Volunteer Computing Platforms.}
	\label{tab:desirable_features}
\begin{tabular}{|l|l|}
\hline
\rowcolor[HTML]{F5F5F0} 
\textbf{Accesibility}                                                        & \scriptsize Users must connect to the platform easily                                                                                   \\

\rowcolor[HTML]{FFFFFF} 
\textbf{\begin{tabular}[c]{@{}l@{}} Adaptability /\\ Dinamicity\end{tabular}} & \begin{tabular}[c]{@{}l@{}}\scriptsize The environment is ever-changing, devices are connected \\ \scriptsize and disconnected at will\end{tabular} \\
\rowcolor[HTML]{F5F5F0} 
\textbf{Availability}                                                        & \begin{tabular}[c]{@{}l@{}}\scriptsize The platform should be available regardless of any \\ \scriptsize problem\end{tabular}                       \\
\rowcolor[HTML]{FFFFFF} 
\textbf{Fault Tolerance}                                                     & \begin{tabular}[c]{@{}l@{}}\scriptsize The platform should be tolerant of failures and \\ \scriptsize disconnections\end{tabular}                    \\
\rowcolor[HTML]{F5F5F0} 
\textbf{Heterogenity}                                                        & \begin{tabular}[c]{@{}l@{}}\scriptsize All connected devices could have different performance,\\ \scriptsize hardware, and OS\end{tabular}          \\
\rowcolor[HTML]{FFFFFF} 
\textbf{Programmability}                                                     & \begin{tabular}[c]{@{}l@{}}\scriptsize Developers should be able to add applications to the\\ \scriptsize platform quickly\end{tabular}             \\
\rowcolor[HTML]{F5F5F0} 
\textbf{Scalability}                                                         & \begin{tabular}[c]{@{}l@{}}\scriptsize The platform must handle a growing amount of\\ \scriptsize connections \end{tabular}                          \\
\rowcolor[HTML]{FFFFFF} 
\textbf{Security}                                                            & \begin{tabular}[c]{@{}l@{}}\scriptsize The machines of the volunteers should not be \\ \scriptsize compromised \end{tabular}                          \\
\rowcolor[HTML]{F5F5F0} 
\textbf{Usability}                                                           & \scriptsize The platform has to be easy to deploy and use                                                                               \\ \hline
\end{tabular}
\end{table}

\subsection{JSDoop Manifest}\label{sec:advantages_jsdoop}

In this section, we define the requirements for JSDoop in terms of the desirable features for VC systems (Table~\ref{tab:desirable_features}). 

\emph{Accesibility.} In JSDoop, the only requirement to participate as a volunteer is to have a modern web browser. Therefore, any device that has one can connect to the platform in a simple and accessible way.

\emph{Adaptability/Dinamicity.} Our proposed system dynamically adapts to the number of connected devices no matter how many or who are connected (as long as the QueueServer supports the number of connections). JSDoop distributes tasks to anyone who requests them. Also, it does not mark a task as completed until it receives an explicit acknowledgment of its completion. Moreover, the Initiator can set a maximum time to solve a task. Then, if a task is not resolved within the maximum time, it is added back to the pending queue. In the meantime, JSDoop allows a dynamical number of devices joining or leaving the system at will (just by closing the web page).

\emph{Availability.} The web browser has become ubiquitous in our daily lives without the need to install any software and can be found in desktop computers, computing centres and mobile phones. Also, the QueueServer is able to recover from failures without losing execution status.

\emph{Fault Tolerance.} JSDoop recovers from failures easily. If a volunteer disconnects while solving a task, the task is added back to the queue. Also, there is a maximum time to solve a task, so if a volunteer freezes during the resolution of a task, the task is added back to the queue.

\emph{Heterogenity.} JSDoop is cross-platform. Any device can collaborate by simply connecting to a URL through a browser.

\emph{Programmability.} JSDoop uses Node.js, which is an open-source, cross-platform JavaScript run-time environment. JavaScript is the most popular language of 2019 according to the latest data from StackOverflow (\url{https://insights.stackoverflow.com/survey/2019}). In the same survey, Node.js appears as the most commonly used in the category of \emph{Other Frameworks, Libraries, and Tools}.

\emph{Scalability.} Depending on the problem being solved, JSDoop can offer a very high scalability. For example, it is possible to use several QueueServers in which each one stores a different type of task. At the same time, it is possible to use a distributed DataServer.

\emph{Security.} Web Browser runs web pages in a sandbox. Sandboxing is an important security technique that isolates programs, preventing malicious or malfunctioning programs from damaging your computer. Of course, in this proof of concept, there are many factors that are not being taken into account and should be taken very seriously in the final version of the library. In future versions of JSDoop, security measures will be implemented to protect volunteers, servers and information. In this paper, we are not going to go into detail on how to get all the necessary security in a VC system because we think it would be beyond the scope of this work.

\emph{Usability.} JSDoop does not require the installation of additional software. A volunteer joins collaboration just opening a URL in the web browser.

\section{Distributed Deep Learning}\label{sec:distributed_deeplearning}

In this section, we discuss distributed client-side deep learning. In particular, we explore the main issues that are pushing deep learning into the client-side (inference and training). Next, we present the classic method of distributed training of NNs and the problems associated with it, such as bottleneck bandwidth and resource-constrained devices. Subsequently, we comment on the solutions proposed to tackle these problems. Later, we briefly review the state-of-the-art of large-scale distributed training of NNs on resource-constrained devices. And finally, we reveal why we believe that this is a particularly exciting area to continue exploring.

Nowadays, mobile data is collected from heterogeneous sources and stored in multiple distributed data centres. With the increase of data volume, it is impractical to move all mobile data to a central data centre to run deep learning applications~\cite{shi2016edge}. Nonetheless, privacy risk and violation make it prohibited to transfer the data of individuals directly to third parties~\cite{wang2018deep}. Many data owners (e.g., medical institutions, insurance companies, ...) cannot share data for reasons of privacy and confidentiality. Others do not want to share their privacy (eg., personal photos or voice records)~\cite{shokri2015privacy}. Therefore, researchers are exploring the possibility of moving the inference and training of NNs to web browsers~\cite{ma2019moving} and mobile devices~\cite{chen2019deep}, and analysing the feasibility of deploying deep models through geographically distributed servers and simultaneously constrained devices, with high efficiency and low overhead, maintaining user privacy and avoiding the problem of sending all data to the cloud~\cite{zhang2019deep, bonawitz2019towards}.

There are many challenges for large-scale distributed training of a NN. The classic optimizer for NN training, stochastic gradient descent (SGD), require low-latency and high-throughput connections to the training data. In parallel distributed data, each machine has a copy of the complete model and calculates the gradients with local mini-batches, and the parameters or gradients of the local model are frequently synchronized (synchronously or asynchronously) to achieve a global consensus of the model learned. This method requires a large bandwidth to synchronize the gradients between all machines producing a bottleneck~\cite{li2014communication}. Another problem is that the training and even inference of NNs is a pretty heavy process, and although the computing power of end devices (e.g., mobile devices) is continuously growing, they are still machines with constrained resources, and some DL tasks remain a heavy workload for them~\cite{ma2019moving, chen2019deep}.

There are many solutions which are out the scope of this paper to reduce the problem of bottlenecks in gradients synchronization~\cite{jiang2018linear}, and the problem of training NNs in resource-constrained devices~\cite{wang2018deep}. A common method for large-scale training is to increase the size of the mini-batch~\cite{krizhevsky2012imagenet}. However, this solution increases computational complexity and memory requirements, so it is not recommended for constrained devices. Another solution is to compress the gradients. \textit{Aji et al.}~\cite{aji2017sparse} proposes the sparsification method, which consists in sending only the absolute values larger than a threshold. They managed to reduce communication and achieved a speedup of 22\% on 4GPUs. \textit{Lin et al.}~\cite{lin2017deep} propose gradient sparsification with momentum correction and local gradient clipping, achieving a gradient compression ration from 270x to 600x for a wide variety of CNNs and RNNs without losing accuracy. \textit{Wen et al.}~\cite{wen2017terngrad} propose the quantization method which consists in quantizing the gradients to three numerical levels \{-1, 0, 1\} reducing the communication cost with none or little accuracy lost on image classification. There are also methods for quantizing the entire model, including gradients~\cite{de2015taming, han2015deep}, which can be useful for deep learning on resource-constrained devices~\cite{meng2017two, zhang2019deep}. There are also solutions for distributing the NN across computing hierarchies, between the cloud, the edge and end devices~\cite{chen2019deep, teerapittayanon2017distributed}.

Some of the latest works on large-scale distributed training of NNs are paying full attention to data privacy and avoiding sending all data to the cloud for processing. In 2015, \textit{Shokri et al.}~\cite{shokri2015privacy} designed a distributed SGD algorithm for parallelized and asynchronously training of a NN using data from different sources. In this model, participants train independently on their datasets and selectively share small subsets of their key parameters during training. In 2016, \textit{Gaia et al.}~\cite{hsieh2017gaia} study a distributed NN training in the scenario where devices are geographically distributed across a large area decoupling the synchronization within a data centre (LANs) from the synchronization across different data centres (WANs). In 2016, \textit{McMahan et al.}~\cite{mcmahan2016communication}, a Google researcher, proposes an asynchronous method to train the NN called \textit{Federated Learning} leaving the data distributed on the mobile devices and learns a shared model by aggregating locally computed updates getting a reduction in required communication rounds by 10 to 100x as compared to synchronized SGD. In 2017, Google adopted Federated Learning~\cite{mcmahan2017federated} (https://www.tensorflow.org/federated/federated\_learning), which allows mobile phones to collaboratively learn a shared prediction model while keeping all training data on the device, decoupling the ability to do machine learning from the need to store the data in the cloud.

Although these techniques are very useful in reducing the problems of large-scale distributed training of a NN, we do not want to divert attention from our work to the way the NN is trained. In future work, we plan to implement some of these techniques to achieve greater scalability. However, in this paper, we want to show that with technologies such as WASM, TensorFlow.js, and JSDoop, among others, we can use the same processing power of end devices through web browsers as with native applications. We can even use the GPU of these devices for applications such as distributed neural network training. The training of a NN is not the final objective of this paper, however, we believe it was necessary to show that it is possible to train NNs using the web browser in a distributed way because we think it is a very interesting area of research that we intend to deepen in future work and that we also believe can inspire other researchers to develop new architectures and adapted models.

In this section, we have shown that large-scale distributed training of a NN using devices with limited resources is not an easy task. However, we confirmed how many researchers are exploring this possibility for reasons as important as privacy and the enormous amount of data generated every day for all devices connected to the Internet.

\section{JSDoop}\label{sec:jsdoop}

As we have seen in previous sections, there are currently many initiatives to implement VC platforms. We have shown that the biggest problems of the traditional VC are accessibility, usability and security (Table~\ref{tab:desirable_features}). Some of the installations of traditional VC applications are not simple. People do not want to install new programs because they want to protect their privacy, out of laziness, fear of spyware or lack of knowledge. On the other hand, as previously explained, BBVC solves all these problems. Also, the performance of web browser applications is approaching native programs.  \par

Therefore, we propose JSDoop\footnote{Code available in \url{https://github.com/jsdoop/}}, a library for distributed collaborative HPC in the web browser in which organizations or individuals (\textbf{Initiator}) create new collaborative projects to be solved by volunteers using their browsers. In other words, an HPC BBVC system. 

Our design and implementation were guided by the desirable features (Table~\ref{tab:desirable_features}). However, we must bear in mind that we are creating a proof-of-concept and not a final product. Therefore, in the final product, some features like security must have greater importance.\par

\begin{figure}[!h]
    \centering
	\includegraphics[width=\columnwidth]{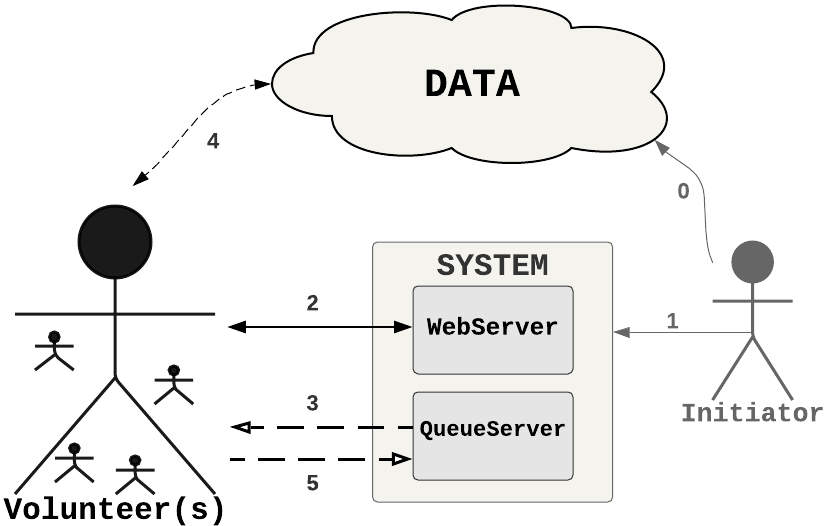}
	\caption{JSDoop Use Diagram.}
	\label{fig:system_diagram}
\end{figure}

Figure~\ref{fig:system_diagram} depicts JSDoop at a high level. An initiator setups a problem using our library. Then, JSDoop divides the problem into smaller tasks that are stored in different queues and uploads the code that solves the tasks in a web server. Later on, volunteers open the web page hosted by the \textbf{WebServer} and start processing the tasks until completion. Note that no installation is required, i.e., the code is running in the background of the web browser. 

The following subsections explain our proposal, as well as the execution flow in more detail.
In our design, we can distinguish the next elements: \textbf{Initiator}, \textbf{WebServer}, \textbf{QueueServer}, \textbf{DataServer}, and \textbf{Volunteers}.

\subsection{Volunteer}
A \textbf{Volunteer} \textit{is anyone who connects to the project URL} without worrying about how the problem is being solved. Because the connection is made through a web browser, the volunteer can use almost any type of device, be working on other tasks at once, and even surfing the Internet using other browser windows. \par 

When the volunteer does not want to continue collaborating, he/she can simply close the browser. If no one is collaborating, the problem simply stops its execution because the QueueServer only manages queues, not the execution of task resolution.\par

\subsection{Initiator}

The \textbf{Initiator} \textit{is a person or an organization that creates a voluntary collaborative project using JSDoop}. 
The Initiator has to configure how to divide the problem into smaller tasks (e.g., map and reduce). It has to provide the URL of the QueueServer and the DataServer. Finally, the Initiator must implement the code that is dependent on the problem to be solved, that is, how each type of task into which that problem has been divided is resolved. From then on, the Initiator does not participate again in the solution of the problem.

\subsection{WebServer}

The \textbf{WebServer} \textit{stores the HTML and JavaScript code necessary for the program to start} in the volunteer's browser. The URL of the QueueServer, the URL of the DataServer, and the code to solve the different types of tasks (e.g., map and reduce) which decide the flow of execution.\par

\subsection{QueueServer}
The \textbf{QueueServer} \textit{is responsible for storing the different queues of tasks needed to solve a problem}. The volunteer accesses the corresponding queue that indicates the code of the task being solved and will send the results of its execution to another results queue. To achieve good load balancing there can be multiple QueueServer in which each stores a different type of queue.

\subsection{DataServer}

There are many well-known solutions on the market for implementing a database server such as Microsoft Azure SQL, IBM Db2, MongoDB Atlas, Redis, among others. This problem has already been solved. Therefore, \textit{JSDoop does not care about the type of }\textbf{DataServer}\textit{ implementation. JSDoop just needs to know where the data is and how it can be accessed.}

\subsection{Flow of Execution}
In this section, we describe the flow of execution of JSDoop step by step from scratch. We use the \textbf{message queue pattern} (Figure~\ref{fig:system_diagram}).\par

\textbf{Step 0.} The initiator configures the DataServer and makes it available over the Internet using any of the well-known solutions on the market. We are using Redis.\par

\textbf{Step 1.} The initiator specifies how the problem will be divided into tasks and develops the code for each type of task (e.g., map and reduce). This code is specific to each type of problem and is the only code that must be added by the initiator so that JSDoop knows how to solve it. Also, the initiator provides the URL of the QueueServer and the DataServer. Then, JSDoop uploads the tasks to different specialized queues depending on the type of task in the QueueServer, and uploads the JavaScript or WASM code to the WebServer.\par
\textbf{Step 2.} When a volunteer connects to a URL of a project using the web browser, a program is executed in background in a transparent way. This program contains the flow of execution.\par
\textbf{Step 3.} This background program gets tasks from a queue on a server and solves them one by one. The execution flow decides the queue in QueueServer where to download the next task from. Finally, the result of the execution of a task can be sent to another queue. This is a chained execution flow.\par

\textbf{Step 4.} JSDoop automatically accesses the DataServer when it is neccesary (i.e., CRUD operations).\par

\textbf{Step 5.} JSDoop allows tasks transactions (i.e., ACID properties), ensuring that tasks are not removed from the queue until an ACK is received from all of them. Also, some tasks have to be synchronized. To achieve the synchronization, JSDoop offers two different solutions. In the first solution, when a task completes its execution, it sends a message to a particular queue. Then, the dependent tasks wait until they receive that message. In second one, dependent tasks check if a data has been modified in the DataServer before starting.\par

\begin{figure*}[!h]
    \centering
	\includegraphics[width=\textwidth]{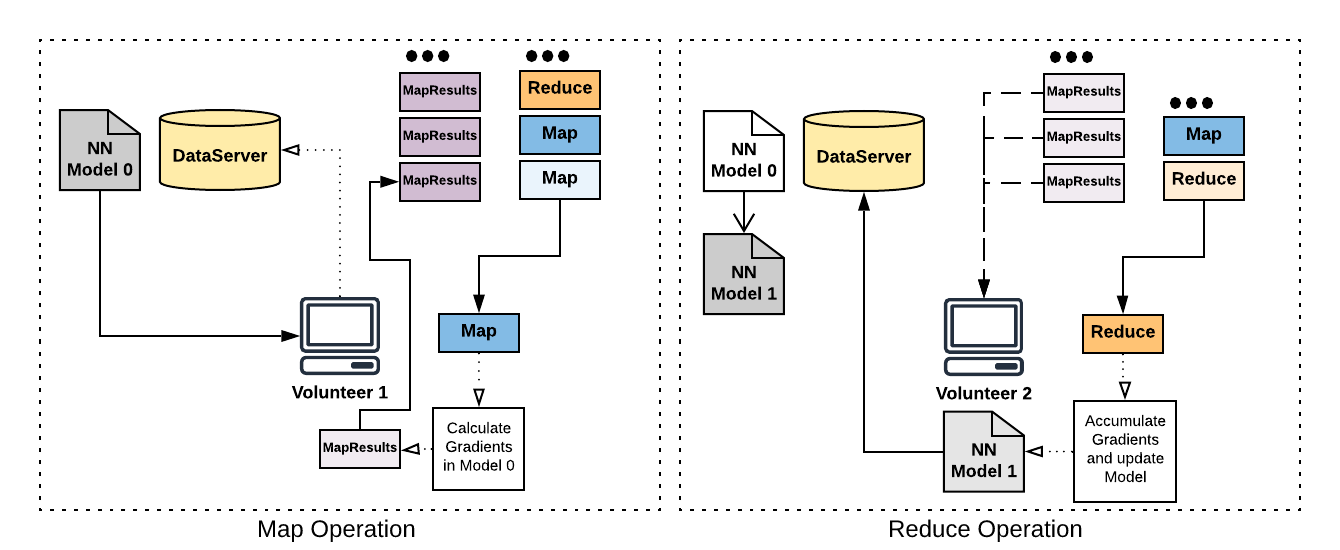}
	\caption{NN Training using Map-Reduce paradigm.}
	\label{fig:map_reduce_training}
\end{figure*}

\subsection{Implementation}

We have implemented JSDoop using NodeJS (version 10.15, \url{http://nodejs.org}). Therefore, we are able to execute a program written using our library in the web browser as well as a native application (e.g., in the console).\par
For communications, in the client side, we use STOMP over WebSocket. The QueueServer is implemented using RabbitMQ (AMQP protocol) (\url{https://www.rabbitmq.com}) for handling the queues. Also, we have implemented a DataServer in Redis (\url{https://redis.io}) to store the data used in the experimentation.\par  
Finally, we implemented a data parallel distributed learning algorithm~\cite{dean2012large, seide20141,strom2015scalable} to train an LSTM-based RNN~\cite{hochreiter1997lstm} to predict the next character, given an input text~\cite{mckeown1992text} using JSDoop and TensorFlow.js. An artificial NN~\cite{goodfellow2016deep} is a model of statistical learning, inspired by biological neural networks. A NN is composed of interconnected neurons, which send messages to each other. Each connection has a weight that can be adjusted systematically according to inputs and outputs. The neurons of an NN are organized into three main parts: the input layer, the hidden layer and the output layer. A NN model is defined by: (i) the number of hidden layers; (ii) the number of nodes in each hidden layer; (iii) how the layers are connected; (iv) which activation function is used; (v) weights on the graph edges. A RNN is a class of NN where connections between nodes form a directed graph along a temporal sequence.\par%

JSDoop is more appropriate for iterative problems because it is possible to create tasks using a loop. In this implementation, we use the map-reduce paradigm~\cite{lammel2008google}. The map-reduce paradigm is based on the following, simple concepts: (i) iteration over the input; (ii) computation of key/value pairs from each piece of input; (iii) grouping of all intermediate values by key; (iv) iteration over the resulting groups; (v) reduction of each group. Map-reduce consists of two main funcionality: map and reduce. Map funcionality take the data and distribute it for computation. Reduce funcionality aggregates the result of computation. This technology is fully successful and integral part of popular systems like Haddop and Spark.\par

In this case, we have as input a text file we want to learn to predict. Also, we have a model of a NN which we want to train. To train the NN, we use backpropagation~\cite{goodfellow2016deep}. First, we initialize the weights of the NN randomly. Second, we do the forward propagation to calculate the actual output. Third, we calculate the loss using categorical cross entropy~\cite{goodfellow2016deep}. Fourth, we use the RMSprop optimizer~\cite{goodfellow2016deep} which deals with the derivative of the loss function.  Fifth, we propagate back the error from the end to the start obtaining the gradients of each weight. Finally, we update the weights.\par

Instead of updating the weights each iteration, we use mini-batch gradient descent~\cite{goodfellow2016deep} which consists in updating the weights after every batch of size \textit{N} accumulating prediction errors. To perform the distributed training (see Figure~\ref{fig:map_reduce_training}), we create a map task to calculate the gradients in each mini-batch. We also create a reduce task that accumulates the calculated gradients and updates the NN model. Both tasks are stored in a \emph{InitialQueue} in the \emph{QueueServer}. When a map task calculates gradients, it sends the results to a \emph{MapResultsQueue}. Before a reduce task is going to accumulate gradients, it downloads all calculated gradients from the \emph{MapResultsQueue}, then it accumulates gradients and updates the NN model in the \emph{DataServer}.\par

The NN model is stored and shared in the DataServer, and it is updated after each reduce task. The NN model has an ID identifying the model version. Each map task has an ID that identifies the version of the model to which the calculation of the gradients is to be made. If the required version is not yet available, the task waits for updating of the NN model by the corresponding reduce task. \par

\section{Experimental Study}\label{sec:experiments}

We implemented a data parallel distributed learning algorithm~\cite{dean2012large, seide20141,strom2015scalable} using JSDoop and Tensorflow.js, and we used our implementation to train an LSTM-based RNN~\cite{hochreiter1997lstm} to predict the next character, given an input text~\cite{mckeown1992text}. The objective of our experimentation is to show that it is possible to train a neural network in a BBVC-based system, as well as to show its scalability.

We divided each data batch into smaller \emph{mini-batches} and computed the gradient of them. Then, we accumulated these mini-batch gradients (to rebuild the original data batch) and applied the gradient to the shared model. To ease the reproducibility, we used an example (sequential) presented in the documentation of TensorFlow.js (\url{https://github.com/tensorflow/tfjs-examples}) as the basis for our implementation.

We studied the performance of our distributed browser-based training using the (relative) speedup and efficiency in two scenarios. First, (\emph{A}) we trained the NN using a cluster of computers. Second, (\emph{B}) we repeated the experiment in a real scenario: a University Classroom. And finally, (\emph{C}) we compared our results to the sequential version (i.e., we used TensorFlow.js on a single browser, without JSDoop) of the same algorithm using the absolute speedup. For a detailed insight on the metrics refer to~\cite{foster1995designing}.

But before we present the experimental results, it is important to remark that in this work we do not intend to improve any distributed neural network training algorithm, nor improve the text generation state-of-the-art, but to show the feasibility of a distributed training in the web browser. Therefore, we might have selected a different problem, e.g., time series prediction or speech recognition, without altering the \emph{objective}, i.e., training a neural network in distributed web browsers.

\subsection{JSDoop on a Cluster of Computers}

To begin with our experimentation, we defined an RNN architecture consisting of two fully connected stacked layers of 50 LSTM cells each, and a dense softmax activation output layer. Then, we trained the network using the parameters presented in Table~\ref{table:training-params}, RMSprop optimizer with categorical cross entropy loss metric, and TensorFlow.js code (compiled, 0.11.7) as the training dataset.

\begin{table}[!h]	
    \caption{Training parameters}    
    \label{table:training-params}
    \centering    
    \begin{tabular}{ lr }
    \hline
    Parameter & Value \\
    \hline
    Batch size & 128 \\
    Examples per epoch & 2048 \\
    Learning rate & 0.1 \\
    Number of epochs & 5 \\
    Sample length & 40 \\
    \hline
    \end{tabular}
\end{table}

The selection of the values of the parameters (Table~\ref{table:training-params}) and the dataset was based on the default values of TensorFlow.js~\cite{smilkov2019tensorflow} example used as the basis thus we ease the comparison of our results against the sequential version of the training algorithm. We use a \emph{batch size} of 128, which means that 128 samples from the training dataset will be used to estimate the error gradient before the model weights are updated. Each sample has a size of 40 characters. We have 2048 samples in each epoch, i.e., 16 batches of size 128 in each epoch. Finally, we run the algorithm for 5 epochs using a learning rate of 0.1 in each experiment.

Table~\ref{table:distributed-params} presents the data parallel distributed parameters. In this case, each 128 \emph{batch size} is divided into 16 8-size mini-batches. Note that the number of \emph{mini-batch to accumulate} multiplied by the \emph{mini-batch size} is equal to the \emph{batch size}~(Table~\ref{table:training-params}). Moreover, we have a map task for each mini-batch, and a reduce task after all mini-batches add up to one batch, i.e., 16 map tasks and one reduce task per batch. As in the sequential algorithm, the weights are not updated until the size 128 batch is accumulated.

\begin{table}[!h]	
    \caption{Distributed training parameters}    
    \label{table:distributed-params}
    \centering    
    \begin{tabular}{ lr }
    \hline
    Parameter & Value \\
    \hline
    Mini-batch size & 8 \\
    Mini-batch to accumulate & 16 \\
    \hline
    \end{tabular}
\end{table}

We tested six cases: 1, 2, 4, 8, 16, and 32 workers. More specifically, we used a \emph{QueueServer} to host the queues and the data, an \emph{initiator} machine (to enqueue the \emph{tasks}), and a cluster with more than 32 heterogeneous computers of different performances administrated with HTCondor~\cite{thain2005distributed} to run the workers. Only the computers that belong to the cluster \emph{solved} the tasks, and all the computers used in the experimentation were interconnected in an Ethernet LAN. 

To have a controlled scenario, we execute only the JavaScript code of each worker. In other words, instead of opening an HTML that runs the code in a web browser, we ran the code directly using NodeJS. Nonetheless, the JavaScript code runs seamlessly in a modern web browser, as it is shown in the next section.

\begin{figure}[!h]
    \centering
	\includegraphics[width=\columnwidth]{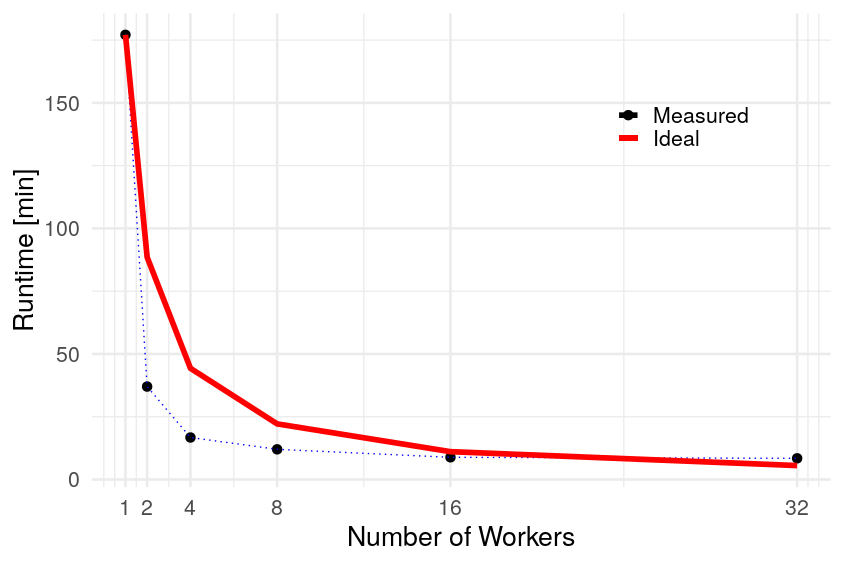}
	\caption{Runtime on a cluster of computers.}
	\label{fig:runtime}
\end{figure}

Figure~\ref{fig:runtime} introduces the parallel runtime for each case tested (\emph{measured}), i.e., the time that elapses from the moment that the first worker starts to the moment that the last worker finishes execution, and the ideal runtime (i.e., linear runtime) considering the number of workers and the runtime of a single worker scenario (solid red line).

\begin{figure}[!h]
    \centering
	\includegraphics[width=\columnwidth]{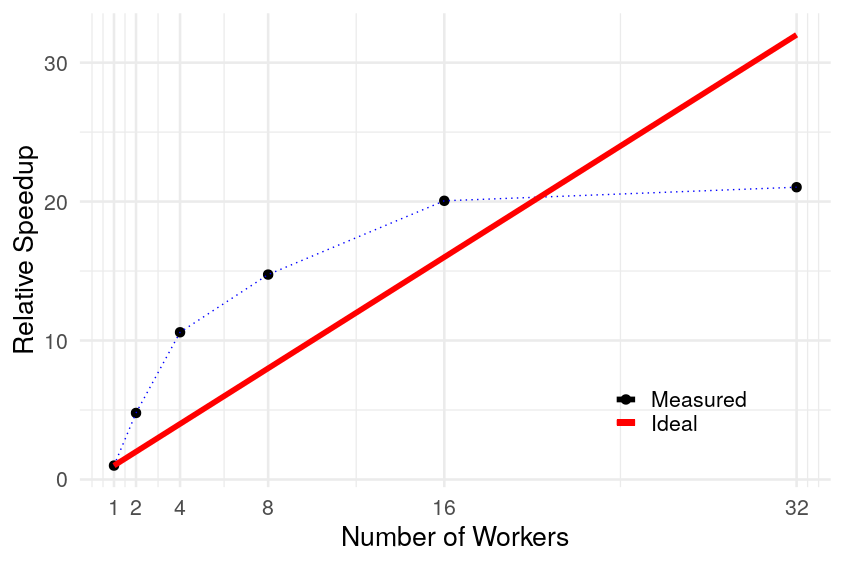}
	\caption{Relative speedup on a cluster of computers.}
	\label{fig:relative-speedup}
	\vspace{-0.8em}
\end{figure}

Figure~\ref{fig:relative-speedup} shows the relative speedup measured, i.e., the speedup ratio calculated using the runtime of the (distributed) algorithm executed in one worker as the reference. Also, the figure shows the ideal speedup (solid red line).

The relative speedup is superlinear for 2, 4, 8, and 16 workers (see Figure~\ref{fig:relative-speedup}). This can be a cache effect. When a problem is executed on a greater number of processors, more of its data can be placed in fast memory. As a result, total computation time will tend to decrease~\cite{foster1995designing}. This is a remarkable result, showing that the problem is more scalable than predicted. On the other hand, as expected because the number of \emph{batches to accumulate} is equal to 16, the speedup is sublinear in the case of 32 workers. Workers need to synchronize after 16 maps with one reduce. Thus, in this scenario, no scalability with more than 16 devices is possible.

\begin{figure}[!h]
    \centering
	\includegraphics[width=\columnwidth]{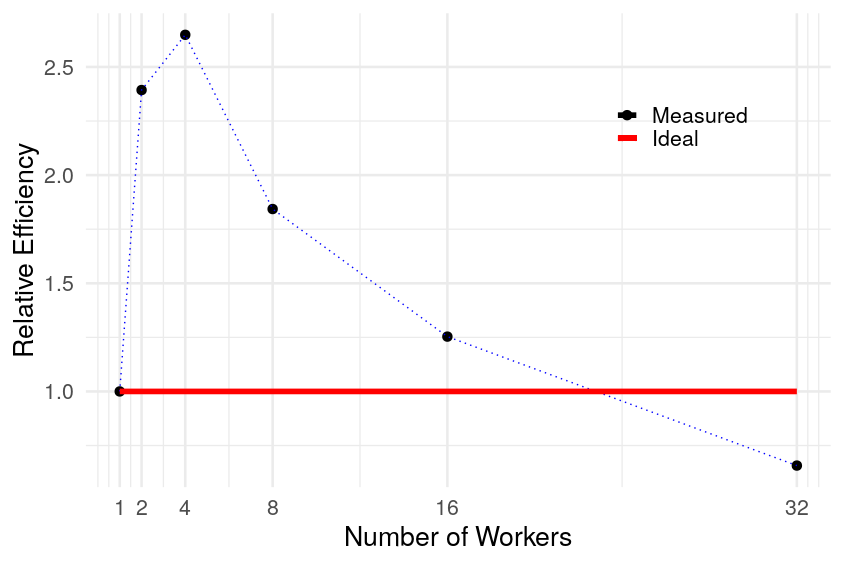}
	\caption{Relative efficiency on a cluster of computers.}
	\label{fig:relative-efficiency}
\end{figure}

Figure~\ref{fig:relative-efficiency} presents the relative efficiency (calculated using the runtime of one worker as the reference) and the ideal efficiency (solid red line). Efficiency is the ratio between the speedup and the number of devices. The ideal efficiency is always equal to one, regardless of the number of workers, i.e., we ideally expect a linear speedup. It is another way of visualizing the same thing we explained above. In line with the previous results, the efficiency is greater than 1 for 2, 4, 8, and 16 workers. Also, it is less than one for 32 workers (because of synchronization).

\subsection{JSDoop in the Classroom}

To continue with our experimentation, we tested our proposal in a real-world scenario: a University Classroom. We solved the same problem defined above, using the same parameters defined in Table~\ref{table:training-params} and~\ref{table:distributed-params}, but using \emph{volunteers} who run web browsers to solve the problem instead of a cluster of computers.

We deployed JSDoop in a web server (Apache HTTP Server) and shared the hyperlink to a group of students asking for their help. To accomplish this task, they only needed to open the hyperlink using a web browser, and once the link was open, they automatically started to contribute to solving the problem.

First, (1) we used 32 volunteer computers to open the link from scratch. Therefore, volunteers were not connected at the same time, but gradually connected (i.e., async-start). Second, (2) once the first problem was solved, we repeated the experiment using the 32 computers already connected to the web page. So all 32 volunteers were connected at the same time (i.e., sync-start). And at last, (3) we asked 16 volunteers to close their web browsers, and then we repeated the experiment with the remaining 16 volunteers. Thus, 16 volunteers were connected at the same time before the problem started (i.e., sync-start).

Table~\ref{table:comparison} summarizes the result of the experiments, where JSDoop-classroom stands for the volunteer-based experimentation and the following number relate to one of the three scenarios described above, and JSDoop-cluster stands for the cluster-based experiments. The runtime is presented in minutes. The best time of each experiment is highlighted in bold. TFJS-Sequential will be explained in the next section.

\begin{table}[!h]	
 
    \caption{Distributed and sequential training}      
    \label{table:comparison}
    \centering    
    \begin{tabular}{ lrrrr }
    \hline
    System & Workers & Runtime & Loss \\
    \hline
    JSDoop-cluster    & 1  & 177.1  & 4.6 \\
    JSDoop-cluster    & 2  &  37.0  & 4.6 \\  
    JSDoop-cluster    & 4  &  16.7  & 4.6 \\
    JSDoop-cluster    & 8  &  12.0  & 4.6 \\
    JSDoop-cluster    & 16 &   8.8  & 4.6 \\
    \textbf{JSDoop-cluster}    & \textbf{32} &   \textbf{8.4}  & \textbf{4.6} \\
    JSDoop-classroom-sync-start   & 16  &   5.4  & 4.6 \\
    \textbf{JSDoop-classroom-sync-start}   & \textbf{32}  &   \textbf{2.5}  & \textbf{4.6} \\
    JSDoop-classroom-async-start   & 32  &   2.7  & 4.6 \\
    \textbf{TFJS-Sequential-128}      & \textbf{1}  &   \textbf{0.9}  & \textbf{4.6} \\
    TFJS-Sequential-8        & 1  &  21.7  & 12.7 \\
    \hline
    \end{tabular}
\end{table}

The results show a significant runtime improvement in the volunteer-based experiment (JSDoop-classroom) compared to the cluster-based one (JSDoop-cluster). This is something expected because voluntary machines and cluster machines have different performances. What is remarkable is that the speedup is similar.  Therefore, we see that we get the same speedup regardless of whether the algorithm runs on the console or on web browsers. On the other hand, the loss (error) of the final trained model is the same in all cases (volunteer and cluster-based experiments). This was to be expected as we have executed the experiment under the same conditions, i.e., the same initial model and an identical order of the data batches.

\begin{figure}[!h]
    \centering
	\includegraphics[width=\columnwidth]{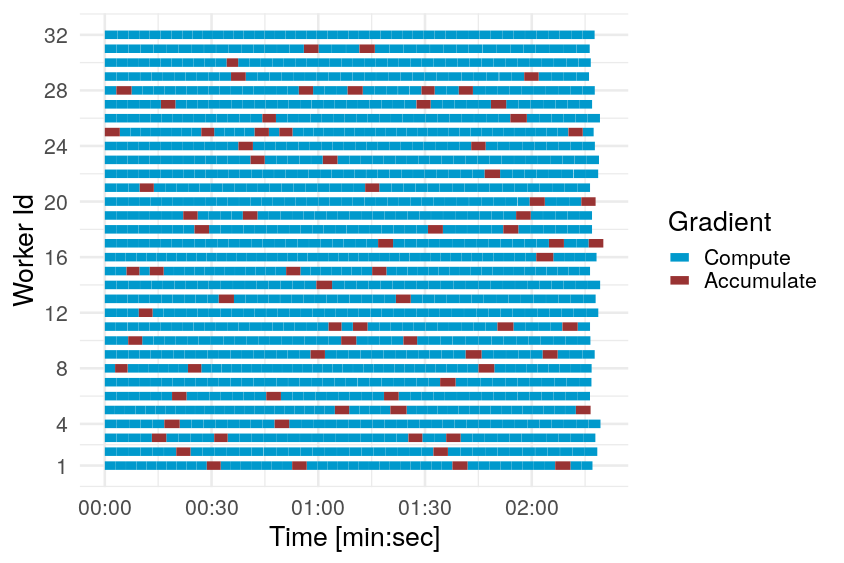}
	\caption{Timeline of JSDoop-classroom-sync-start with 32 volunteers.}
	\label{fig:timeline-16}
\end{figure}

Figure~\ref{fig:timeline-16} depicts the timeline of the experiment conducted with 32 volunteer computers (JSDoop-classroom-sync-start). Volunteers are presented on the \textit{y}-axis, while the \textit{x}-axis represents the time (relative to the start of the test, in minutes). For each volunteer, the figure shows the time spent in processing tasks, i.e., computing the gradient (\emph{Compute}) and accumulating the gradient (\emph{Accumulate}). Note that for each task the timeline is plotted from the moment that a task is received to the time the task is completed. Also, note that all volunteers start at the same time (i.e., sync-start) and that tasks (e.g., Accumulate) are evenly distributed.

\subsection{JSDoop vs. Sequential TensorFlow.js}
Finally, we trained the RNN using the sequential version of the training algorithm (i.e., we used TensorFlow.js on a single browser, without JSDoop) and the parameters defined in Table~\ref{table:training-params}. 

First, we perform an experiment using a batch size of 128 (TFJS-Sequential-128). Moreover, to have a broader perspective of the results, we decided to perform an additional experiment. We repeated the original experiment but using a batch size equal to 8 (TFJS-Sequential-8). In this sense, the sequential and the distributed algorithms compute the gradient the same number of times (Table~\ref{table:distributed-params}, mini-batch size). Therefore, we will compare both approaches under \emph{similar} conditions, i.e., computing and accumulating the same number of times the gradient. Nonetheless, we acknowledge that the optimization problem is slightly different (i.e., we are moving toward the direction of a gradient computed using a smaller data batch), so the loss is expected to be different too.

Table~\ref{table:comparison} summarizes the results for all the experiments. The runtime is presented in minutes. To compare the sequential algorithm with the distributed algorithm, we use the absolute speedup. We define absolute speedup using as the baseline the uniprocessor time of the sequential algorithm~\cite{foster1995designing}.

\begin{figure}[!h]
    \centering
	\includegraphics[width=\columnwidth]{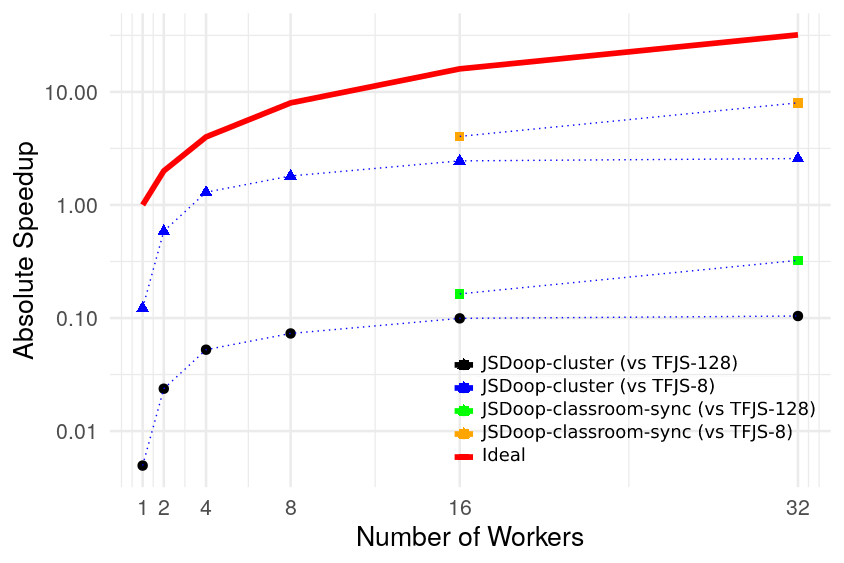}
	\caption{Absolute speedup.}
	\label{fig:absolute-speedup}
\end{figure}

Figure~\ref{fig:absolute-speedup} shows the absolute speedup, calculated using the TFJS-Sequential-128 and TFJS-Sequential-8 results as the runtime reference respectively (refer to Table~\ref{table:comparison}). Note that \emph{JSDoop-cluster (vs. TFJS-128)} corresponds to the speedup calculated for the experiment JSDoop-cluster using TFJS-128 as the runtime reference, and so on. Ideal speedup (i.e., linear absolute speedup) is showed as a solid red line.

Despite that absolute speedups are sublinear, the runtime of the distributed training with 16 volunteers (or more) is close to the TFJS-Sequential-128 runtime. Moreover, the runtime of TFJS-Sequential-8 is larger than most distributed runtimes. As an example, the distributed version running on 32 volunteer computers (JSDoop-classroom-sync-start) is nearly nine times faster than TFJS-Sequential-8. We must bear in mind that the sequential algorithm does not have the synchronization overhead between the nodes over the network. Moreover, the size of the problem is not very large. It is natural that the larger the size of the problem, the greater the benefit of the distributed algorithm. In any case, the results of the experiments show how the algorithm is highly scalable despite the need for synchronization between the nodes.

\section{Threats to Validity}\label{sec:validity}

This section discusses the main threats to the validity of the proof of concept conducted, with special attention to issues that were not reviewed/understood in this study. We can distinguish three main threats: DataServer communication overhead, QueueServer communication overhead and dynamism. Most of them are common in distributed computing systems~\cite{tanenbaum2007distributed}.

Among these weaknesses, maybe the most important is the communication overhead, a very common issue in distributed systems. From the very foundations of JSDoop, where workers are heterogeneous and do not have an \emph{a priori} knowledge of the problem, i.e., they do not have installed any software (besides the web browser), nor the data, all the required information (code, parameters, and data) to solve the problem has to be shared thru the web.

On the one hand, in the design of our experiments, all workers had to synchronize the NN model with the DataServer (i.e., parameter server in NN training). In Section~\ref{sec:distributed_deeplearning}, we explain the solutions offered by the literature to solve the problem of communication overhead in NN training. On the other hand, from a systems architecture point of view, we must distribute the database to mitigate the overhead of several nodes trying to obtain data from the same point. Notice that there are multiple alternatives to mitigate this issue~\cite{ozsu2011principles}. Also, JSDoop uses AMQP, which allows users to distribute the tasks in multiple queues to avoid the bottleneck on the QueueServer. A different server can host each queue, and we can use a load balancer to choose the correct queue. Load balancing is common on distributed platforms.

Regarding dynamism, in this study, we did not analyze the overhead caused by devices dynamically joining and leaving the system. The reason we do not perform a dynamism analysis is that we believe this problem deserves a specific work in which we analyze how a task should be divided into subtasks to minimize the total execution time taking into account that tasks can fail and be restarted because of disconnection. This problem is more related to fault tolerance~\cite{wolter2010stochastic}, and it needs a separate paper for it. We can visualize the possibility of a device disconnects during a task resolution as a failure rate of a task in a fault-tolerant system. If a task fails, it must start from scratch. Therefore, we must find a balance between a large task size to avoid communication overhead, while at the same time avoiding a too large task size that causes a high risk due to the failure rate. This problem is common in volunteer computing platforms, and we want to face it more deeply in future research. There are many proposals to tackle this issue, ranging from dynamic task scheduling~\cite{page2005dynamic} to the dynamic adaptation of the master-worker paradigm~\cite{andre2009dynamic} or to retaining volunteers~\cite{darch2010retaining}. Nonetheless, this is still an open issue. Thus we will need further experiments to study this matter.

As a summary, we believe that the balance between the advantages and drawbacks is in favor of JSDoop. Moreover, the results of this study present a solid argument in favor of our goals. JSDoop is an HPC BBVC library that allows us to train a neural network in distributed web browsers.

\section{Conclusions and Future Work}\label{sec:conclusions}

In this study, we introduced JSDoop, a web browser-based volunteer distributed computing system. It can be used by individuals or organizations to implement HPC programs (subject to web browser constraints,) that run in the web browser in a distributed way, allowing volunteers to collaborate by simply accessing a URL.

We have conducted a proof-of-concept using our proposed system and TensorFlow.js to train an RNN that generates text (i.e., a complex and topical problem). We have tested the performance of our proposal using a cluster of 32 heterogeneous computers and in a real-world scenario, with up to 32 volunteers. Also, we compared our results to the state-of-the-art web browser-based neural network training.

The results presented in this study show that web browser-based distributed neural network training is feasible. Furthermore, JSDoop proved to be an adequate implementation of BBVC achieving high scalability and allowing to add/delete volunteers dynamically during execution, without losing information (tasks). Also, the training of a NN is just one of the many applications that JSDoop can does. JSDoop is a general purpose HPC BBVC library with which you can solve a wide variety of different problems and, in this work, we have been able to adapt the training of a NN to JSDoop.

As future work, we propose to study the task assignment management based on the performance of heterogeneous devices (e.g., PCs, mobile phones, tablets, Raspberry Pi, among others) and to study the performance of the system running on a large group of highly heterogeneous devices. Also, we plan to implement JSDoop as a \emph{Platform as a Service} (PaaS), allowing individuals and organizations to upload their problems (code) into the platform, to get a link to share with volunteers that will solve the problem.

Finally, we want to remark the importance of finding proper ways of motivating the volunteers to participate in this type of efforts.  In this sense, in the future, we propose to explore gamification as a way of attracting volunteers.

\balance

\bibliographystyle{IEEEtran}
\bibliography{main}

\EOD

\end{document}